\documentclass[twocolumn,10pt,superscriptaddress, amsmath,amssymb, aps]{revtex4-1}
\usepackage{graphicx}% Include figure files
\usepackage{dcolumn,todonotes}% Align table columns on decimal point
\usepackage{bm}% bold math
\usepackage{braket}
\usepackage{xcolor}
\usepackage{units,ulem}
\usepackage{bbold}
\usepackage{multirow}

\newcommand{\beq}{\begin{equation}}
\newcommand{\eeq}{\end{equation}}

\renewcommand{\emph}{\textit}

\begin{document}

\title{Three-observer Bell inequality violation on a two-qubit entangled state}

\author{Matteo Schiavon}
\affiliation{Department of Information Engineering, University of Padova, I-35131 Padova, Italy}
\author{Luca Calderaro}
\affiliation{Department of Information Engineering, University of Padova, I-35131 Padova, Italy}
\author{Mirko Pittaluga}
\affiliation{Department of Information Engineering, University of Padova, I-35131 Padova, Italy}
\author{Giuseppe Vallone}
\affiliation{Department of Information Engineering, University of Padova, I-35131 Padova, Italy}
\affiliation{Istituto di Fotonica e Nanotecnologie, CNR, Padova, Italy}
\author{Paolo Villoresi}
\affiliation{Department of Information Engineering, University of Padova, I-35131 Padova, Italy}
\affiliation{Istituto di Fotonica e Nanotecnologie, CNR, Padova, Italy}

\begin{abstract}
Bipartite Bell inequalities can be simultaneously violated by two different pairs of observers when weak measurements and signaling is employed. Here we experimentally demonstrate the violation of two simultaneous CHSH inequalities by exploiting a two-photon polarization maximally entangled state. Our results demonstrate that large double violation is experimentally achievable. Our demonstration may have impact for Quantum Key Distribution or certification of Quantum Random Number generators based on weak measurements.
\end{abstract}

\maketitle

\textit{Introduction -} 
The correlations between the costituents of a system characterized by quantum entanglement may reveal  ``the characteristic trait'' of Quantum Mechanics, in the words of one of its founders Erwin Sch\"odinger~\cite{schr35pro}.  This property has been exploited to investigate the very nature of physical reality, with the observation of the violation of  Bell inequalities~\cite{brun14rmp}. 
In its essence,  in a bipartite system  the most
known Bell inequality is the so called Clauser-Horse-Shimony-Holt (CHSH) inequality, based on  dichotomic measurements. In this case, two observers, usually called Alice and Bob, perform independent measurements on their subsystem. They choose randomly between two different measurements $\mathcal A$ or $\mathcal A'$ for Alice and $\mathcal B$ or $\mathcal B'$ for Bob. The CHSH inequality is written as $I_{\rm CHSH}\equiv\langle \mathcal A\otimes \mathcal B\rangle+
\langle \mathcal A'\otimes \mathcal B\rangle+
\langle \mathcal A\otimes \mathcal B'\rangle-
\langle \mathcal A'\otimes \mathcal B'\rangle\leq 2$.
Bipartite entangled states may violate such inequality: in particular, for a two-qubit maximally entangled state, the Bell parameter $I_{\rm CHSH}$ may reach the Tsirelson's bound $2\sqrt{2}$~\cite{cire80lmp}.
The violation of the CHSH inequality certifies the presence of entanglement and rules out the 
possibility of describing quantum physics with a local hidden variable model~\cite{bell64phy}, as recently demonstrated experimentally~\cite{hens15nat,gius15prl,
shal15prl}.

An intriguing property of quantum entanglement is its {\it 	monogamy}~\cite{coff00pra}: given a tripartite state $\rho_{AB_1B_2}$, the larger is the entanglement between two observers, the lower is the entanglement of the third observer with any of the other two. 
A similar monogamy argument holds for ``non-local-realistic'' correlations~\cite{masa06pra,tone09prsa}, whose presence is associated with the violation of Bell inequalities. 
Indeed, given three observers (Alice, Bob1 and Bob2) and assuming non signaling, it is impossible to have a simultaneous
violation between Alice-Bob1 and Alice-Bob2. 

However, as realized in~\cite{silv15prl}, this restriction no longer holds if the non-signaling hypothesis is dropped. Therefore, it is possible to violate the CHSH inequality between two different pairs of observers by using a single two-qubit entangled state and allowing the state received by Bob2 to be first measured by Bob1.
In this case, the state received by Bob2 is dependent on Bob1's basis choice and therefore there is signaling between Bob1 and Bob2.
However, the two Bobs do not have to agree on a common measurement strategy and can in principle be unaware of each other's presence, so they may be considered as independent.
This  no longer holds for more than two observers on the Bob's side: they would have to agree on a measurement strategy, that is, they cannot perform unbiased measurements in order to violate all together the CHSH inequality with Alice~\cite{mal16qph}.

Here we introduce a model to test the limit of the monogamy in the case of a photonic bipartite entangled state, by varying the strength of the weak measure performed by Bob1 in the presence of an independent strong measure realized by Bob2. We aim to observe that the correlations that both Bobs have with Alice may experimentally violate the corresponding CHSH inequalities.  The crucial ingredient to achieve such
double violation is the weak measurement 
performed by one of the observers.
Again, the underlying concept that we exploited here is that by performing the weak measurement at Bob1, a lower 
amount of information is obtained about the system with respect to a   projective, or strong,  measurement. However, such less information is compensated by a lower degree of disturbance on the measured state.

\begin{figure}[t]
    \centering
    \includegraphics[width=\linewidth]{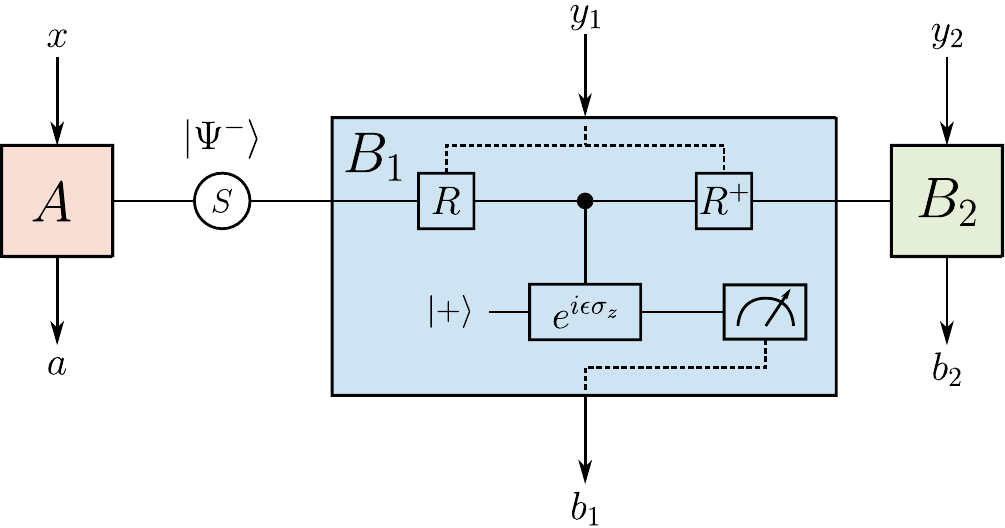}
    \caption{Circuit model of the scheme used for the double violation of the CHSH inequality.}
    \label{fig:circuit}
\end{figure}

\begin{figure*}[t]
\centering
\includegraphics[width=2\columnwidth, clip]{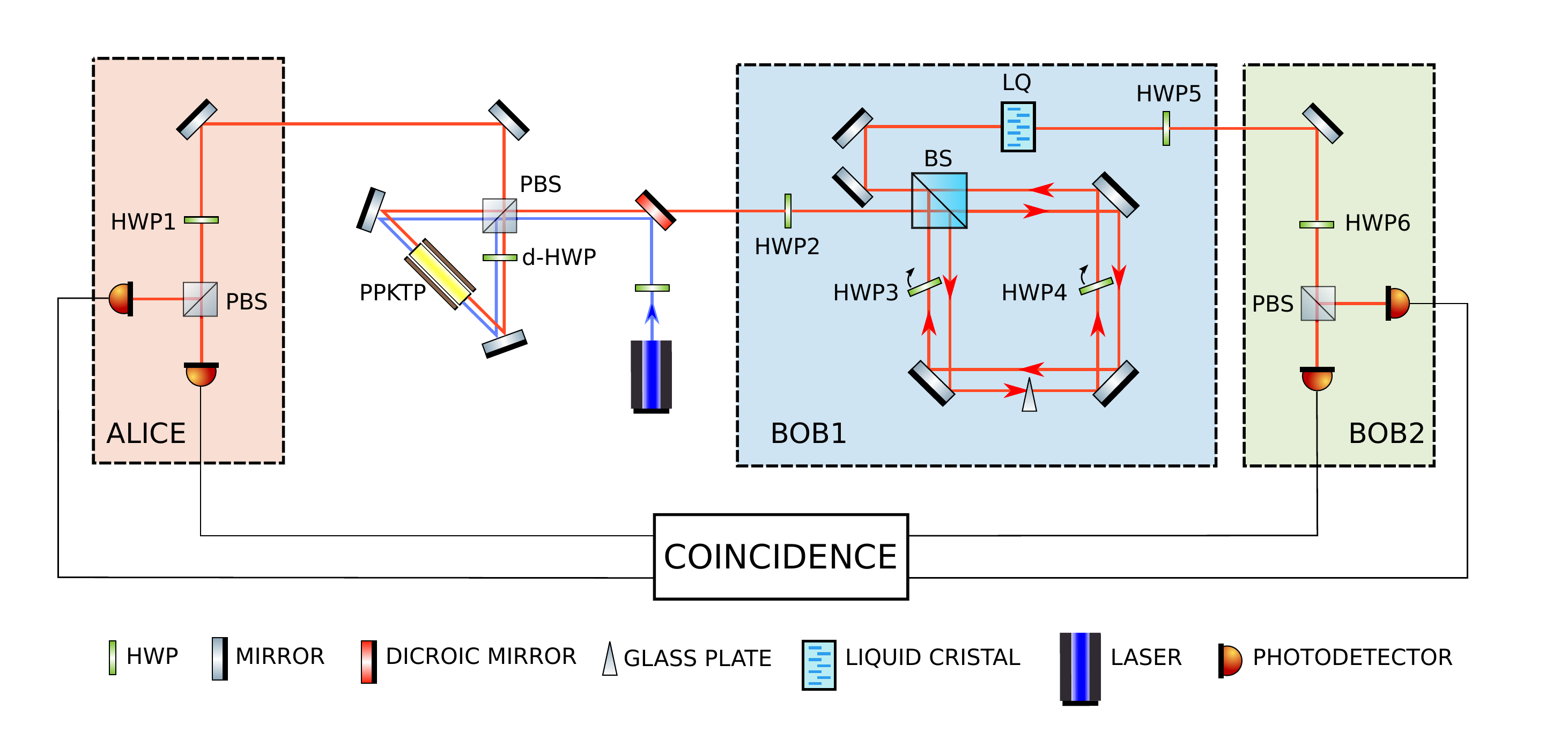}
\caption{Scheme of the experimental setup. The polarization-entangled photon-pair source comprises a PPKTP crystal, in a Sagnac interferometer, pumped by a laser diode at $404.5$ nm. The entangled photons are collected and sent to Alice and Bobs apparatuses. Alice and Bob2 implement a scheme, consisting of a HWP (HWP1 and HWP6) and a PBS, to measure the polarization on two linear bases. The transmitted and reflected photons from the PBS are detected by single photon avalanche diodes. Bob1's apparatus performs the weak measurement. HWP2 and HWP5  implement the transformations $R$ and $R^\dagger$, respectively. HWP3 and HWP4 are placed in a sagnac interferometer with clockwise and anticlockwise paths spatially separated. In particular, HWP3 (HWP4) is placed in the clockwise (anticlockwise) path, and is used as a phase retarder between horizontal and vertical polarization. The phase difference between the two paths is adjusted by tilting a thin glass plate. Finally, a liquid crystal is used as a phase retarder between horizontal and vertical polarization.}
\label{fig:Setup}
\end{figure*}
\textit{Theoretical model -} 
The scheme allowing the double violation of the CHSH inequality is presented in Fig. \ref{fig:circuit}. Three  observers, Alice ($A$), Bob1 ($B1$) and Bob2 ($B2$) perform
some measurements on a two-particle entangled
state $\ket{\Psi^-}$.
In particular, each of the observer choose independently between two different dichotomic
observables by using a single random bit
(respectively denoted by $x$, $y_1$ and $y_2$
for the three observers) taking values in $0$ 
or $1$. The binary outcomes ($\pm$)
of the dichotomic measurements are respectively given by $a$, $b_1$ and $b_2$. By measuring the probability of the outcomes, two CHSH parameters $I_{\rm CHSH}$ can be evaluated as:
\beq
\begin{aligned}
I^{(1)}_{\rm CHSH}&=\sum_{x,y_1}
(-1)^{x\cdot y_1}[p(a=b_1|x,y_1)-p(a\neq b_1|x,y_1)],
\\
I^{(2)}_{\rm CHSH}&=\sum_{x,y_2}
(-1)^{x\cdot y_2}[p(a=b_2|x,y_2)-p(a\neq b_2|x,y_2)].
\end{aligned}
\eeq
In order to violate both  CHSH inequalities
$I^{(j)}_{\rm CHSH}\leq 2$ it is necessary
that Bob1 performs a weak measurement on his subsystem: as said, such weak measurement cannot extract full information from the state, but it will also not completely disturb it.

More precisely, a weak measurement scheme consists of entangling the system under measurement with an ancillary system and then strongly measuring the ancilla.
The measurement scheme of Bob1 exploits a controlled phase gate ${\rm CP}_\epsilon=
\ket{H}\bra{H}\otimes\openone+
\ket{V}\bra{V}\otimes e^{i\epsilon\sigma_z}$
to entangle an ancillary qubit, prepared in the state $\ket{+}$, with the system coming from the source. The ancilla qubit is then measured in the $\left\{ \ket{+},\ket{-} \right\}$ basis. Depending on 
the amount of rotation $\epsilon$ in the controlled phase gate it is possible to vary the strength of the measurement. For instance
by setting $\epsilon=\pi/2$ correspond to a strong projective measurement in the 
$\{\ket H,\ket V\}$ basis.
The above scheme implements a controllable-strength measurement of the system in the $\left\{ \ket{H},\ket{V} \right\}$ basis.
In order to generalize the measurement to an arbitrary basis $\left\{ \ket{\omega_{y_1}},\ket{\omega_{y_1}^\perp} \right\}$ it is necessary to rotate the state with a rotation matrix $R_{y_1}$ such that $R_{y_1} \ket{\omega_{y_1}} = \ket{H}$ and $R_{y_1} \ket{\omega_{y_1}^\perp} = \ket{V}$. 

The complete scheme of the three-observer Bell violation is shown in Fig. \ref{fig:circuit}.
The source $S$ prepares the state in the singlet state
\begin{equation}
    \ket{\Psi^{-}} = \frac{\ket{H}\ket{V} - \ket{V}\ket{H}}{\sqrt{2}},
\end{equation}
and sends one state to Alice and the other one to the Bobs.
Alice performs a strong measurement on her system in the basis $\left\{ \ket{u_x}, \ket{u_x^\perp} \right\}$, according to her measurement choice $x \in \{0,1\}$, obtaining a result $a \in \{+,-\}$.
After her measurement, the state at the Bobs' side is projected into $\ket{u_x^{(-a)}}$, where $\ket{u_x^{(+)}} \equiv \ket{u_x}$ and $\ket{u_x^{(-)}} \equiv \ket{u_x^\perp}$.
The state at the entrance of Bob1 system $\ket{\psi_{a|x}} = \ket{u_x^{(-a)}}$ can be rewritten in his measurement basis as $\alpha \ket{\omega_{y_1}} + \beta \ket{\omega_{y_1}^\perp}$, where $\alpha = \braket{\omega_{y_1}|u_x^{(-a)}}$ and $\beta = \braket{\omega_{y_1}^\perp|u_x^{(-a)}}$.
Inside Bob1 measurement apparatus, the joint state $\ket{\psi_{a|x}} \otimes \ket{+}$ is transformed by the rotation into $\left( \alpha \ket{H} + \beta \ket{V} \right) \ket{+}$.
The controlled phase gate implements the unitary $e^{i\epsilon\sigma_z}$ on the ancilla qubit conditioned on the system qubit being $\ket{V}$, transforming the state into
\begin{equation}
\label{Bob1out}    \alpha \ket{H} \ket{+} + \beta \ket{V} \left( \cos\epsilon \ket{+} + i\sin\epsilon\ket{-} \right). 
\end{equation}
After the inverse rotation $R_{y_1}^\dagger$, the joint state becomes
\begin{equation}
    \ket{\psi_{a|xy_1}} = \alpha \ket{\omega_{y_1}} \ket{+} + \beta \ket{\omega_{y_1}^\perp}  \left( \cos\epsilon \ket{+} + i\sin\epsilon\ket{-} \right),
\end{equation}
or, in the density matrix formalism, $\rho_{a|xy_1} = \ket{\psi_{a|xy_1}}\bra{\psi_{a|xy_1}}$.
Bob1 measures the second qubit of this state in the basis $\left\{ \ket{+},\ket{-} \right\}$, while Bob2 performs a strong measurement on the first qubit in the basis $\left\{ \ket{v_{y_2}},\ket{v_{y_2}^\perp} \right\}$, dependent on his choice of $y_2 \in \{0,1\}$.
They both obtain a separate joint probability distribution with Alice
\begin{align}
    p(a,b_1|x,y_1) &= {\rm Tr}\left[ \left( \openone \otimes \Pi_{b_1} \right) \rho_{a|xy_1} \right] p(a|x), 
    \\
    \notag
    p(a,b_2|x,y_2) &= \sum_{y_1} p(y_1) {\rm Tr}\left[ \left( \Pi^{y_2}_{b_2} \otimes \openone \right) \rho_{a|xy_1} \right] p(a|x),
\end{align}
where $\Pi_{b_1}$ and $\Pi^{y_2}_{b_2}$ are, respectively, the projectors to Bob1 and Bob2 measurement states.

If Alice chooses to  measure in the directions $-(Z+X)/\sqrt{2}$ or $(-Z+X)/\sqrt{2}$ and the Bobs choose to measure in the $Z$ or $X$ directions (corresponding to the rotation matrices $R_0=\openone$ and $R_1=\frac12(\sigma_z+\sigma_x)$ for Bob1), 
the above 
probabilities predict the following
values of the CHSH parameters:
\begin{align}
I^{(1)}_{\rm CHSH} &= 2\sqrt{2}\sin^2\epsilon, \label{eq:I1CHSH}\\
I^{(2)}_{\rm CHSH} &= \sqrt{2}(1 + \cos\epsilon). \label{eq:I2CHSH}
\end{align}

\textit{Experimental setup -} Fig. \ref{fig:Setup} illustrates the setup of our experiment. Entangled photon pairs are produced by using a 30 mm periodically poled KTP crystal in a polarization-based Sagnac interferometer~\cite{kim06pra} and collected into single-mode fibers. The photons are produced by Spontaneous Parametric Down Conversion from a single mode UV laser at $\unit[404.5]{nm}$ and $\unit[4]{mW}$ of power.
A polarization controller is placed at Alice's side in order to compensate polarization rotations induced by fiber birefringence. 
The receiving apparatuses of Alice and Bob2 implement a scheme for measuring the polarization of the photons. An half wave plate (HWP) is placed before a polarizing beam splitter (PBS) to measure on two arbitrary orthogonal states in the X-Z plane of the Bloch sphere. The transmitted and reflected photons from the PBS are detected using single photon avalanche diodes (SPAD).
Before reaching Bob2, the photons pass through Bob1's apparatus. The polarization rotation $R$ and $R^{\dagger}$ are performed by HWP2 and HWP5 respectively. The control phase gate is implemented by exploiting a Sagnac interferometer, with clockwise and anticlockwise paths spatially separated, plus a liquid crystal (LQ) after one output port. In this configuration the ancilla is given by the paths of the interferometer. HWP3 and HWP4 provide a phase retardation between horizontal and vertical polarizations, say $\epsilon_0$ and $\epsilon_1$ respectively. Indeed, their slow axis are parallel to the vertical polarization and the phase retardation is adjusted by tilting them. The phase difference between the two paths $\phi$ is adjusted by tilting a thin glass plate. Finally, the liquid crystal is used as a phase retarder between horizontal and vertical polarization.

\begin{figure}[t]
\centering
\includegraphics[width=\columnwidth]{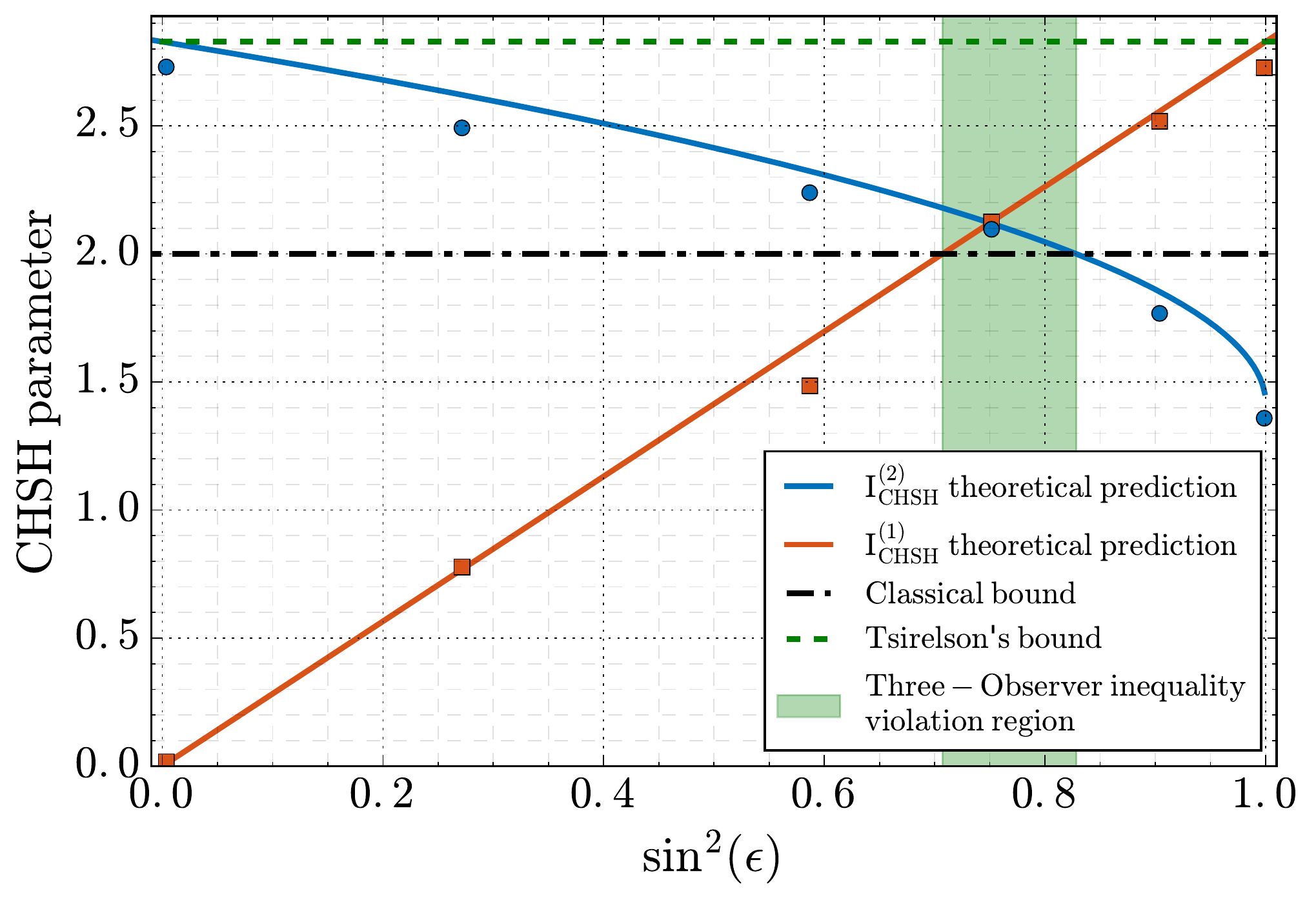}
\caption{Measurements of $I^{(1)}_{\rm CHSH}$ (squares) and $I^{(2)}_{\rm CHSH}$ (diamonds) for several values of $\epsilon$. The red and green solid lines show the expected values of $I^{(1)}_{\rm CHSH}$ and $I^{(2)}_{\rm CHSH}$ (Eqs.~\ref{eq:I1CHSH} and \ref{eq:I2CHSH}), while the dash-dotted and dashed lines indicate classical and Tsirelson's bounds respectively. The green region highlights the values of $\epsilon$ in which double violation is expected. Poissonian errors are within the
dimension of the points.}
\label{fig:CHSHweak}
\end{figure}

To show that Bob1's apparatus implements the scheme in Fig.~\ref{fig:circuit}, suppose a pure state $\ket{\psi_{in}} = \alpha \ket{\omega_{y_1}} + \beta \ket{\omega_{y_1}^{\perp}}$ is given as an input. The HWP2 rotates the state into $\alpha \ket{H} + \beta \ket{V}$, which enters the interferometer. After the first passage through the beam splitter (BS) and the HWP3 and HWP4, the state becomes
\begin{equation}
\frac{i e^{i\phi}}{\sqrt{2}}\left(\alpha\ket{H} + \beta e^{i\epsilon_0}\ket{V}\right) \ket{0} + \frac{1} {\sqrt{2}} \left( \alpha\ket{H} + \beta e^{i\epsilon_1} \ket{V} \right) \ket{1},
\end{equation}
being $\ket{0}$ and $\ket{1}$ the clockwise and anticlockwise path respectively and $\phi$ the phase difference between them. At the exit of the interferometer we have that, setting the glass plate such that $\phi=0$, the state is
$\alpha\ket{H} \ket{2} + \beta e^{i\frac{\epsilon_0+\epsilon_1}{2}} \ket{V} \left( \cos\epsilon \ket{2} + \sin\epsilon \ket{3} \right)$,
where $\ket{2}$ and $\ket{3}$ are the two output ports of the interferometer and $\epsilon = \frac{\epsilon_1 - \epsilon_0}{2}$. If the phase retardation of the LQ \footnote{The reasoning holds considering a LQ after each output port of the interferometer. In our case, 
we need a single LQ since we use only one output port.} is set at $-\frac{\epsilon_0+\epsilon_1}{2}$, the state is changed into
$\alpha\ket{H} \ket{2} + \beta \ket{V} \left( \cos\epsilon \ket{2} + \sin\epsilon \ket{3} \right)$.
Hence, the Sagnac interferometer plus the LQ implements a control phase gate with phase $\epsilon = \frac{\epsilon_1-\epsilon_0}{2}$. Finally, the HWP5 rotates the state into
\begin{equation}
\ket{\psi_{out}} = \alpha\ket{\omega_{y_1}} \ket{2} + \beta \ket{\omega_{y_1}^\perp} \left( \cos\epsilon \ket{2} + \sin\epsilon \ket{3} \right).
\end{equation}
In our setup, we look at one single output port at a time, swapping $\ket{2}$ and $\ket{3}$ by tilting the glass plate. Indeed, by changing the global phase $\phi$ from $0$ to $\pi$ the transformation $\ket{2} \rightarrow -\ket{3}$ and $\ket{3} \rightarrow \ket{2}$ applies.
The measurement in the states $\ket{2}$ or
$\ket{3}$ on the above state corresponds
to the measurement in the basis 
$\{\ket+,\ket-\}$ on the state \eqref{Bob1out}.

Alice chooses between the two measurement bases $(X-Z)/\sqrt{2}$ and $-(X+Z)/\sqrt{2}$; these bases are obtained from the HWP1 rotated by the angles $11.25^{\circ}$ and $33.75^{\circ}$ respectively. Bob1 and Bob2 choose between measurement bases $Z$ and $X$, given by a rotation of $0^{\circ}$ and $22.5^{\circ}$ of the couple HWP2-HWP5 and of HWP6, respectively. Furthermore, Bob1 tilts the glass plate into two positions corresponding to the phase difference $\phi = 0$ and $\phi = \pi$, to swap the output port. In total, a single measurement of the Bell parameters $I^{(1)}_{\rm CHSH}$ and $I^{(2)}_{\rm CHSH}$ requires $16$ data acquisitions, one per each different configuration of the HWPs and the glass plate. In all measurements, we set $30$ seconds of data acquisition, with an average coincidence rate of $700$ counts per second. According to the standard procedure in Bell inequality violation we did not subtracted accidental coincidences.

\textit{Results -} We measured $I^{(1)}_{\rm CHSH}$ and $I^{(2)}_{\rm CHSH}$ for several values of $\epsilon$ in the range $[0,\pi/2]$. 
The procedure used to estimate $\epsilon$ is detailed in appendix.
Fig.~\ref{fig:CHSHweak} shows the
obtained experimental results, demosntrating a good agreement with the theoretical model. For $\epsilon = 0$, there is no interaction between the polarization and the ancillary state. Indeed, $I^{(1)}_{\rm CHSH}$ is comparable to $0$, while $I^{(2)}_{\rm CHSH}$ is close to the Tsirelson's bound. By increasing $\epsilon$, we demonstrate an increase of $I^{(1)}_{\rm CHSH}$ and a reduction of $I^{(2)}_{\rm CHSH}$, following the expected theoretical curves.

\begin{figure}[h]
\centering
\includegraphics[width=\columnwidth]{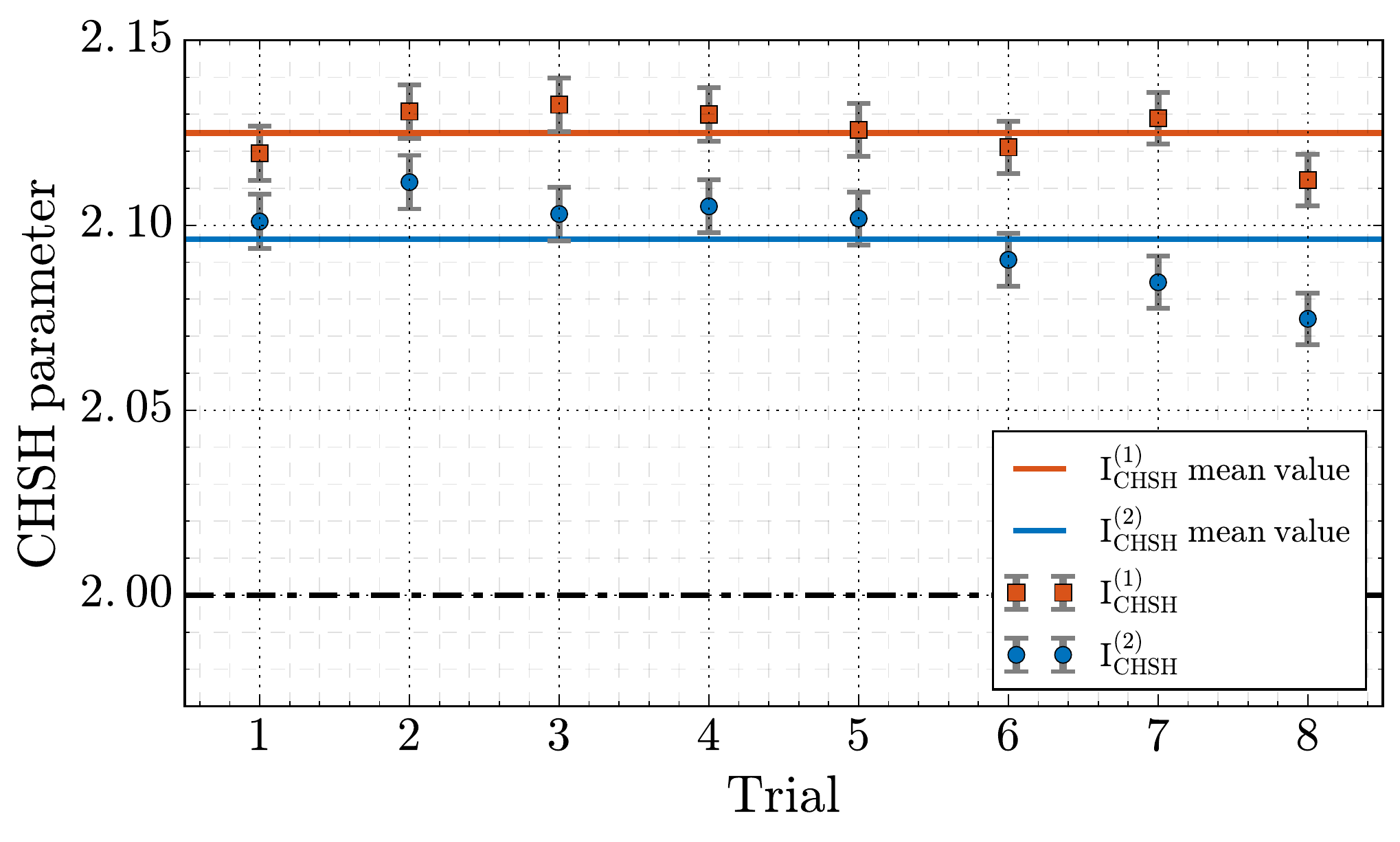}
\includegraphics[width=\columnwidth]{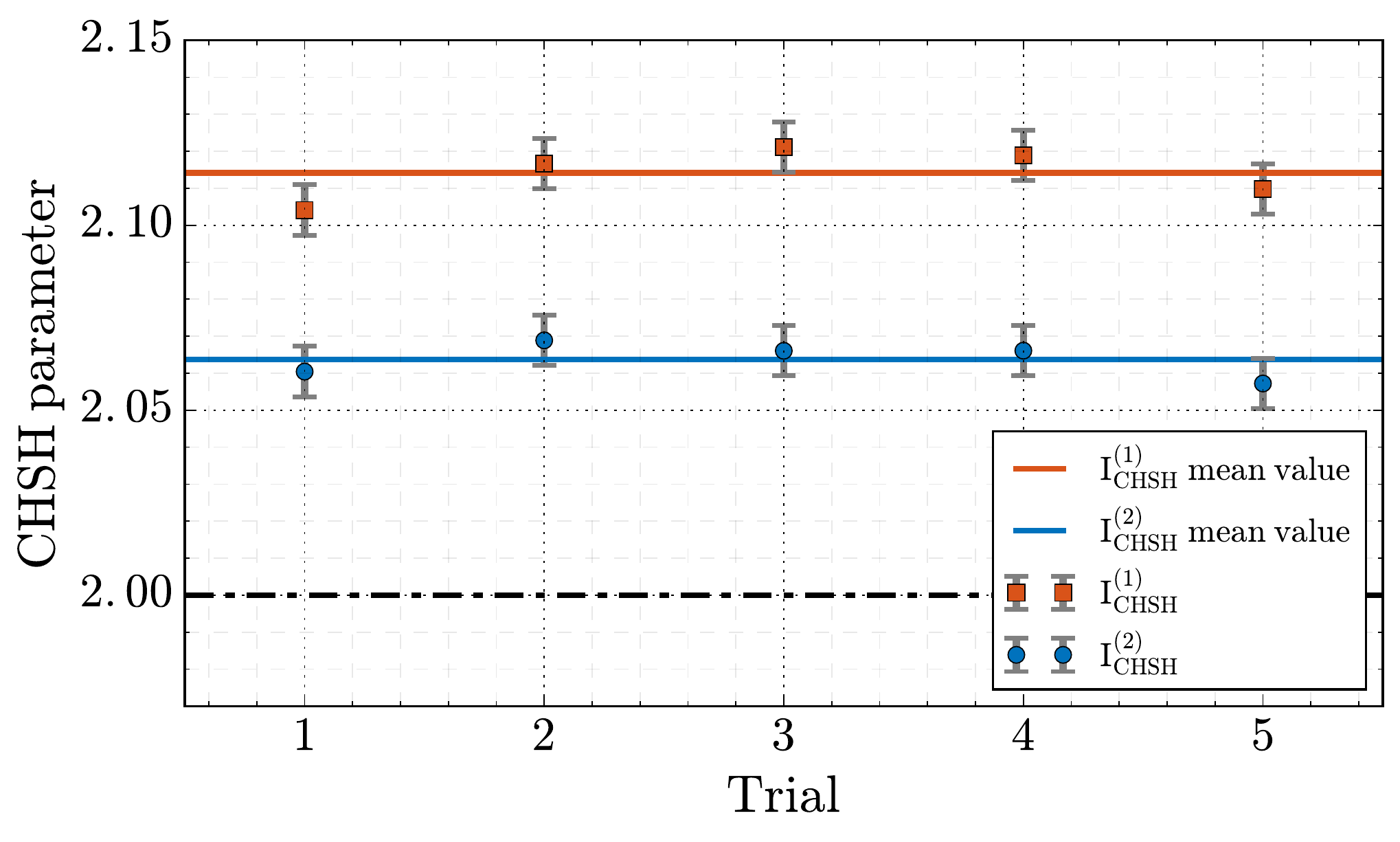}
\caption{Measurements of $I^{(1)}_{\rm CHSH}$ (squares) and $I^{(2)}_{\rm CHSH}$ (diamonds) in two consecutive series of trials. Red and blue solid lines indicate the mean value of $I^{(1)}_{\rm CHSH}$ and $I^{(2)}_{\rm CHSH}$ respectively. \textbf{(Top)} Eight consecutive trials were performed in an hour, with $\epsilon= 1.049 \pm 0.002$. Considering the poissonian error, the measurements show a violation of $10$ standard deviations, fluctuating around mean values of $I^{(1)}_{\rm CHSH}= 2.125 \pm 0.003 $ and $I^{(2)}_{\rm CHSH}= 2.096 \pm 0.003$. \textbf{(Bottom)} Another series of five consecutive trials were performed within a hour, with $\epsilon= 1.053 \pm 0.002$. Similarly to the previous case, all the measurements show a violation of {$10$} standard deviation, fluctuating around mean values of $I^{(1)}_{\rm CHSH} = 2.114 \pm 0.003$ and $I^{(2)}_{\rm CHSH} = 2.064 \pm 0.003$.} \label{fig:DoubleViolation}
\end{figure}

The interesting region is the one around $\epsilon = \pi/3$, where both $I^{(1)}_{\rm CHSH}$ and $I^{(2)}_{\rm CHSH}$ are expected to be above the classical bound -- green region in Fig.~\ref{fig:CHSHweak}. To give a larger statistical evidence of the double violation
in this region, we performed consecutive measurements with two different values of $\epsilon$. In Fig. \ref{fig:DoubleViolation} \textbf{(Top)}, we show the results of $8$ consecutive measurements with $\epsilon = 1.049 \pm 0.002$. In all trials, both $I^{(1)}_{\rm CHSH}$ and $I^{(2)}_{\rm CHSH}$ were above the classical bound, fluctuating around the mean values $I^{(1)}_{\rm CHSH}= 2.125 \pm 0.003 $ and $I^{(2)}_{\rm CHSH}= 2.096 \pm 0.003$. The data acquisition of a single trial took about eight minutes to finish, for a total acquisition time of one hour. This proves the reproducibility of the violation and the stability of our setup. A second series of trials, with $\epsilon = 1.053 \pm 0.002$ is shown in Fig. \ref{fig:DoubleViolation} \textbf{(Bottom)}. Similarly to the previous case, both $I^{(1)}_{\rm CHSH}$ and $I^{(2)}_{\rm CHSH}$ are above the classical bound for the entire period of the acquisition, with $I^{(1)}_{\rm CHSH} = 2.114 \pm 0.003$ and $I^{(2)}_{\rm CHSH} = 2.064 \pm 0.003$.

\textit{Conclusions -} We have shown experimentally that
a double CHSH inequality violation between
two different pairs of observers can be achieved by using a single two-qubit entangled state of two photons. We recall that the three observers choose randomly between the two possible measurements, with no agreement on
the measurement strategy.
Thanks to the  stability of our setup (larger than 1 hour), we could perform several double violations increasing the statistical evidence
of the experiment.
The double violation was tested and repeated for different values of $\epsilon$, the
interaction strength of Bob1's weak
measurement. The experimental data well reproduce the theoretical model when $\epsilon$ is changed.

It is worth noticing that by slightly changing the measurement setting at Bob1's side it is possible to obtain an optimal weak measurement.
Indeed, if the phases of the glass plate are set to $\phi=\phi_0$ and $\phi=\phi_0+\pi$
the value of the $A$-$B1$ inequality can be varied as  
$I^{(1)}_{\rm CHSH} = 2\sqrt{2}[\cos\phi_0-\cos(\phi_0-\epsilon) ])$, maximized to
$I^{(1)}_{\rm CHSH} = 2\sqrt{2}\sin\epsilon$ for $\phi_0=\epsilon-\pi/2$.
The change in $\phi_0$ does not change the value of $I^{(2)}_{\rm CHSH}$: this 
corresponds to a measurement that keeps the
disturbance on the state fixed with a varying information gained on it.
Our scheme demonstrates that even if the weak measurement is not optimal it is possible to  achieve a double violation of the inequality.

The achievement of double violation and the realization of a simple  weak measurement scheme have important applications for Quantum Random Number Generation QRNG, as demonstrated in~\cite{curc15qph},
or for Quantum Key Distribution exploiting weak measurements~\cite{trou15qph}.
In~\cite{curc15qph}, for instance, it was shown 
that, by using sequences of weak measurements to violate a multi-user Bell inequality, it is possible to
certify any amount of random bits from a pair of pure entangled qubits.

{\it Note added - }
While we were completing our work we became aware that a similar manuscript,
showing the experimental violation a double Bell inequality, was recently posted on arXiv~\cite{hu16qph}.

\onecolumngrid
\appendix
\vskip1cm

\begin{center}
\textbf{\large Appendix: Three-observer Bell inequality violation on a two-qubit entangled state}
\end{center}

\textit{Strength of the weak measurement -} In our weak measurement scheme, the strength of the interaction between the state to be measured and the ancilla is parametrized by the phase $\epsilon$. A good estimation of this parameter is needed to have a desirable precision on the measurement strength. In this section we describe the procedure we used to estimate it. 

As pointed out in the \textit{Experimental setup} section, $\epsilon$ depends on the phase retardation $\epsilon_0$ and $\epsilon_1$ given by HWP3 and HWP4, respectively, so that $\epsilon = \frac{\epsilon_1 - \epsilon_0}{2}$. Considering a state $\ket{\psi} = \alpha\ket{H} + \beta\ket{V}$, that enters the interferometer of Bob1, at the output of a single port we have
\begin{equation}
\ket{\psi_{out}}=\alpha\cos\left(\frac{\phi}{2}\right)\ket{H} + \beta e^{i\frac{\epsilon_1+\epsilon_0}{2}} \cos\left(\frac{\phi}{2}-\epsilon
\right) \ket{V},
\end{equation}
where $\phi$ is the phase difference between clockwise and anticlockwise path, and is controlled by the glass plate. In particular, for small angle $\theta$ of the thin glass plate we can omit the refraction and consider the following model for $\phi$
\begin{equation}
\phi(\theta) = \frac{\chi}{\cos\theta} + \phi_0, \qquad \chi = \frac{2\pi}{\lambda}d\Delta n, \label{eq:modelphi}
\end{equation}
being $\theta$ the incident angle of the beam on the plate, $\lambda$ the wavelength, $d$ the thickness of the plate and $\Delta n$ the difference between the refraction indexes of the glass and air. 
The state then passes through the liquid crystal and the HWP5, set with vertical slow axis, adding a phase retardation between $\ket{H}$ and $\ket{V}$. At this point, if Bob2 measures in the $Z$ basis, his outcomes will be
\begin{align}
P_H(\phi) = |\braket{H|\psi_{out}}|^2 &= |\alpha|^2 \cos^2\left(\frac{\phi}{2}\right),\\ 
P_V(\phi) = |\braket{V|\psi_{out}}|^2 &= |\beta|^2 \cos^2\left(\frac{\phi}{2}-\epsilon\right).
\end{align}

\begin{figure}[h]
\centering
\includegraphics[width=0.95\textwidth]{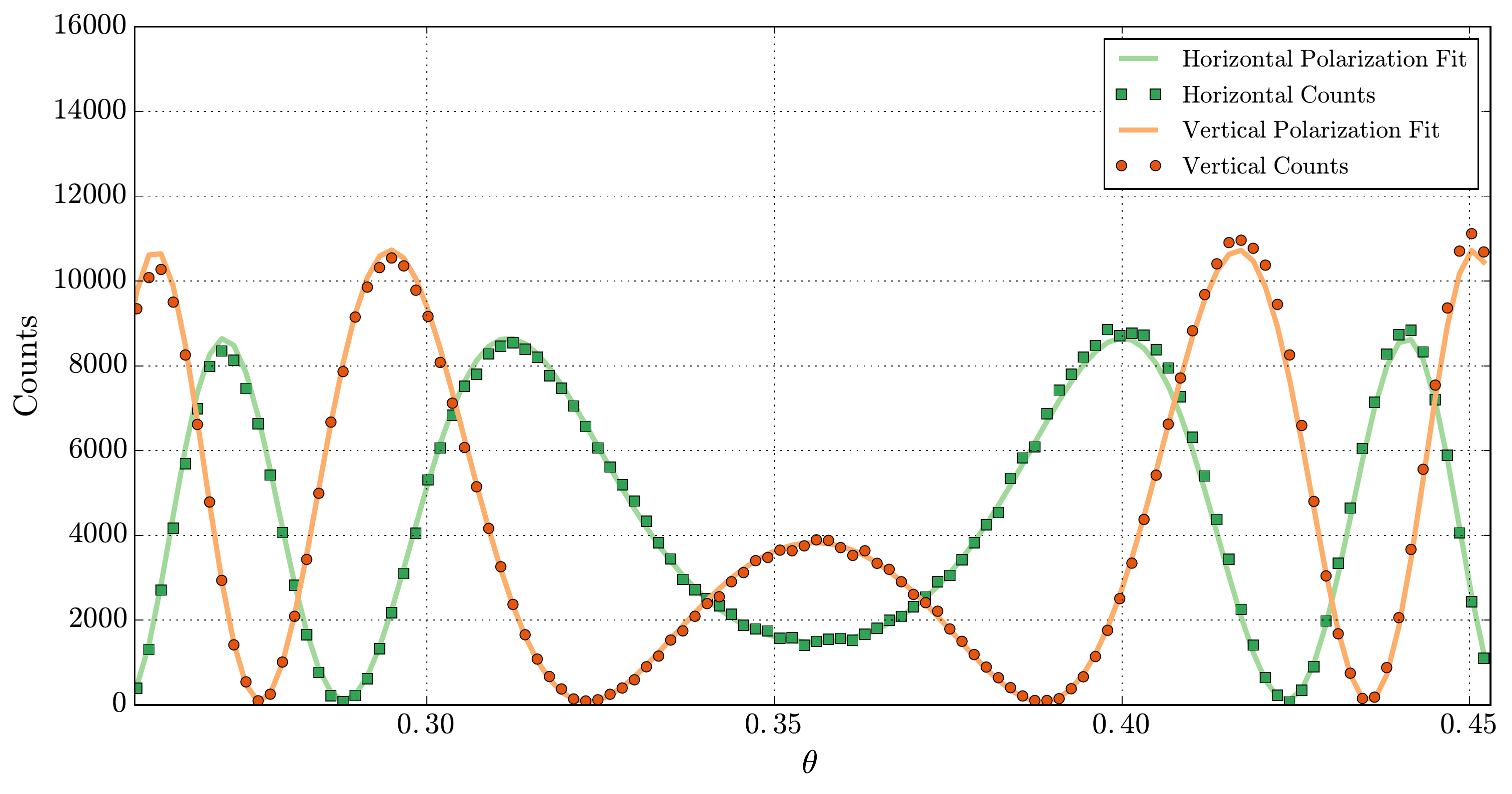}
\caption{Example of the procedure to evaluate the relative phase difference $\epsilon$ given by the interferometer between the horizontal and vertical polarization. In this example the horizontal polarization counts are fitted by the function  ${P_H = 8600\cdot\sin^2(\frac{1185.5}{\cos(\theta-0.356)}+2.45)+98}$, while the vertical polarization  counts are fitted by ${P_V~=~11000\cdot\sin^2(\frac{1185.5}{\cos(\theta - 0 . 356)}+1.40)+68}$. 
$\theta$ is the rotation angle
of the glass plate expressed
in radians.
The phase difference $\epsilon$ is thus $1.049 \pm 0.004$.} \label{fig:epsilon}
\end{figure}

By measuring $P_H$ and $P_V$ for several values of $\theta$ and interpolating with equations
\begin{align}
P_H(\theta) &= I_H \cos^2\left(\frac{\chi}{\cos(\theta - \theta_0)} + \phi_H\right), \label{eq:PH}\\ 
P_V(\theta) &= I_V \cos^2\left(\frac{\chi}{\cos(\theta - \theta_0)} + \phi_V\right). \label{eq:PV}
\end{align}
we can estimate $\epsilon$, since $\epsilon = \phi_H - \phi_V$.
In Fig.~\ref{fig:epsilon}, we show the measurements of Bob2 and the interpolation of Eq.~\ref{eq:PH} and \ref{eq:PV}. The model is in perfect agreement with the experimental data.

\begin{figure}[h]
\centering
\includegraphics[width=0.95\textwidth]{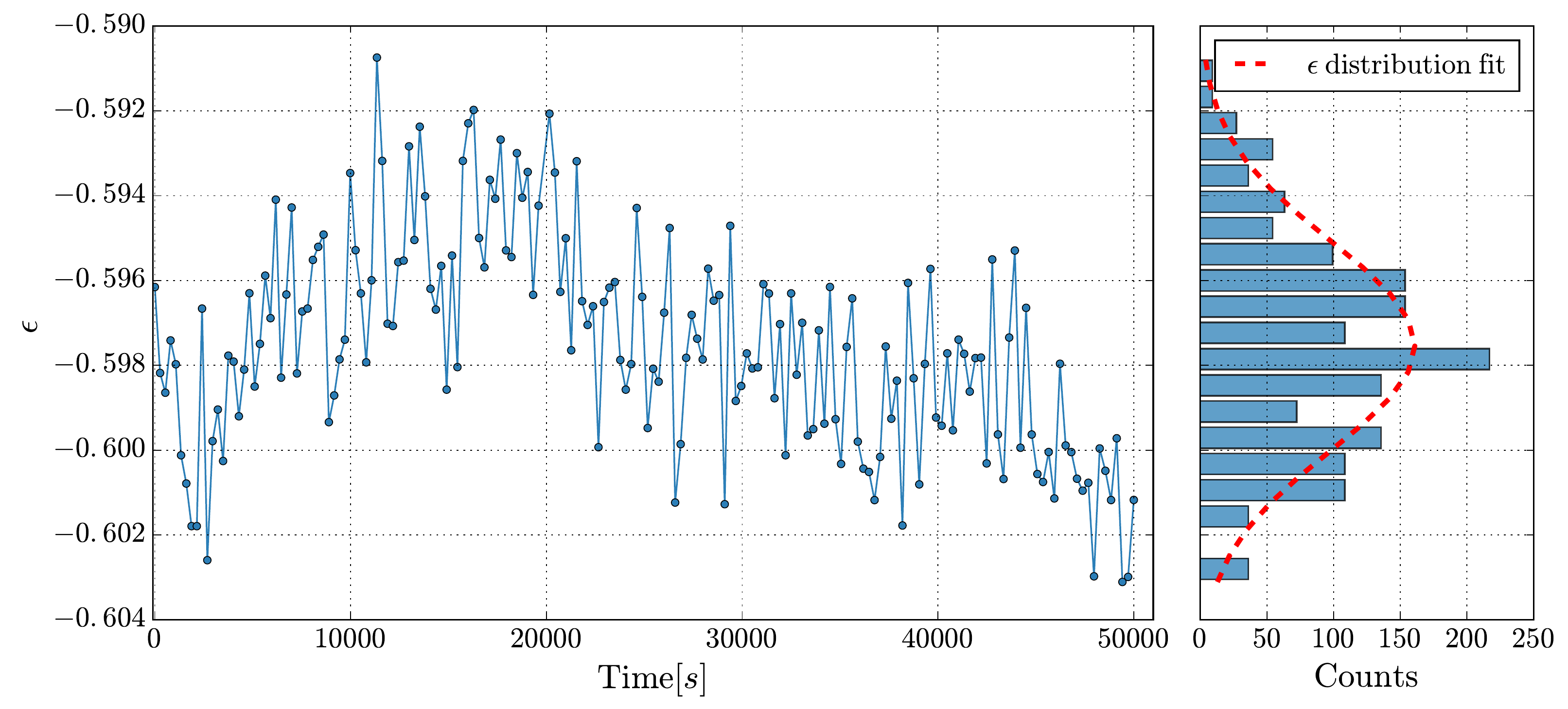}
\caption{Estimation of the $\epsilon$ 
phase displacement over several measurements lasting for a period of thirteen hours. (\textbf{Left}) Each point in the graph represents a different estimation of $\epsilon$  (\textbf{Right}) $\epsilon$ value distribution. The gaussian curve that fits the data is characterized by a mean value $\mu = -0.5975$ and
a standard deviation $\sigma = 0.0025$. 
The standard deviation of the distribution was used to evaluate the error on the estimation of $\epsilon$.} \label{fig:epsilonerror}
\end{figure}

We evaluated the stability of $\epsilon$, by repeating consecutively the analysis above over a period of thirteen hours. Fig.~\ref{fig:epsilonerror} shows that $\epsilon$ remains stable in time. 
The error associated to the single measurement of $\epsilon$ reported in the main text
is the RMS of such thirteen hour measurement.

\end{document}